\documentclass[journal=jacsat,manuscript=article]{achemso}

\usepackage[version=3]{mhchem} 
\usepackage[T1]{fontenc}       
\usepackage{graphicx}

\author{Filip Krzy{\.z}ewski}
 \affiliation {Institute of Physics, Polish Academy of Sciences, Al. Lotnik{\'o}w 32/46,
02-668 Warsaw, Poland}
\email{fkrzy@ifpan.edu.pl}
 \author{Magdalena A. Za{\l}uska--Kotur}
\email{zalum@ifpan.edu.pl}
\affiliation {Institute of Physics, Polish Academy of Sciences, Al. Lotnik{\'o}w 32/46,
02-668 Warsaw, Poland}
\alsoaffiliation{Faculty of Mathematics and Natural Sciences, Card. Stefan Wyszynski University, ul. Dewajtis 5, 01-815 Warsaw, Poland}
\author{Henryk Turski}
\affiliation {Institute of High Pressure Physics Polish Academy of Sciences, Al. Prymasa Tysiaclecia 98, 01-424 Warsaw, Poland}
\author{Marta Sawicka}
\affiliation {Institute of High Pressure Physics Polish Academy of Sciences, Al. Prymasa Tysiaclecia 98, 01-424 Warsaw, Poland}
\author{Czes{\l}aw Skierbiszewski}
\affiliation {Institute of High Pressure Physics Polish Academy of Sciences, Al. Prymasa Tysiaclecia 98, 01-424 Warsaw, Poland}
\title[An \textsf{achemso} demo]
  { Miscut dependent surface evolution   in the process of N-polar GaN$(000\bar 1)$ growth under N-rich condition}

\abbreviations{IR,NMR,UV}

\begin{document}

\begin{tocentry}

\includegraphics[width=9cm]{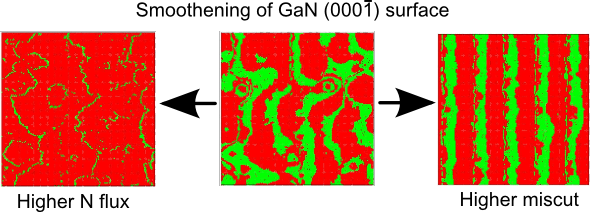} 

\end{tocentry}

\begin{abstract}
 The evolution of surface morphology during the growth of N-polar $(000\bar1)$ GaN under N-rich condition is studied by kinetic Monte Carlo (kMC) simulations for two substrates miscuts $2^o$ and $4^o$. The results are compared with experimentally observed surface morphologies of $(000\bar1)$ GaN layers grown by plasma-assisted molecular beam epitaxy. The proposed kMC two-component model of GaN$(000\bar1)$ surface where both types of atoms: nitrogen and gallium attach the surface and diffuse independently, explains that at relatively high rates of the step flow (miscut angle <$2^o$) the low diffusion of gallium adatoms causes surface instabilities and leads to experimentally observed roughening while for low rates of the step flow (miscut $4^o$), smooth surface can be obtained. In the presence of almost immobile nitrogen atoms under N-rich conditions, the growth is realized by the process of two-dimensional island nucleation and coalescence.  Additionally, we show that higher crystal miscut, lower crystal growth rate or higher temperature results in similar effect of the smoothening of the surface. We show that the surface also smoothens  for the growth conditions with very high N-excess. The presence of large number of nitrogen atoms changes locally  mobility of gallium atoms thus providing easier coalescence of separated  island. 
\end{abstract}

\section{Introduction}
The fact that in wurtzite N-polar GaN, the direction of the polarization fields is reversed compared to the Ga-polar direction, makes the N-polar GaN of interest for a variety of device applications such as high electron mobility transistors, solar cells, sensors and  light emitting diodes \cite{rajan,hoi,keller}. Identification of the optimum growth conditions providing smooth surface morphology N-polar $(000\bar1)$ GaN is one of the key issues to enhance the development of these emerging applications. The growth of atomically flat N-polar $(000\bar1)$ GaN layers by plasma-assisted molecular beam epitaxy (PAMBE) has been proven difficult due to the high adatom diffusion barriers because it is carried out at relatively low temperatures $(750^oC)$. Theoretical works showed that these diffusion barriers are substantially reduced when gallium adlayer is present on the surface during the growth \cite{zywietz,neugebauer,takeuchi,review}. However, previous experimental studies showed that the growth of the N-face GaN must be performed with less than one monolayer (ML) of excess Ga on the growing surface, in contrast to Ga-polarity, when more than 2ML can be stabilized. Therefore, a precise control of the growth conditions is needed to prevent the accumulation of Ga droplets\cite{monroy}. The morphologies reported for N-polar GaN layers grown by PAMBE under N-rich conditions on low miscut substrates were relatively rough either due to the hillocks or pattern formed by interlacing stripes covered by doubly-bunched atomic steps \cite{cheze}. In search for the optimum growth conditions of the device-quality N-polar GaN layers it is essential to understand of the role of surface miscut, growth temperature, growth rate and III/V ratio (N-excess) that may improve surface morphology. It is important to note that in case of metal-organic vapor phase epitaxy (MOCVD), the N-polar $(000\bar1)$ GaN   
films for many years suffered from high surface roughness resulting from the formation of large hexagonal hillocks\cite{zauner}. The use of 4 deg miscut N-polar GaN substrates has proven to be effective method of improving surface morphology \cite{keller}.

In this work the growth of N-polar GaN  $ (000\bar 1)$ surface under N-rich conditions  is investigated theoretically by kinetic Monte Carlo (kMC) simulations and compared to experimental results in order to identify the optimum growth conditions leading to smooth layer morphologies and understand the role of substrate miscut, temperature, growth rate and N-excess.
 We built two-component model of GaN $(000\bar{1})$ surface. We use the model that is based on  the one previously implemented to study growth and sublimation of GaN(0001) surface \cite{krzyz1,krzyz2,krzyz3,krzyz4}. 
The model is developed in such a way that both types of atoms building crystal: nitrogen and gallium are controlled. They attach to the surface,  diffuse and  detach independently. They  can be adsorbed at the steps or form islands by  nucleation at the terraces.  In this work we study the influence of substrate miscut on the morphology of $(000\bar{1})$ GaN layers by comparing  the simulated systems  to  the morphologies of N-polar GaN $000\bar1)$ layers grown bt PAMBE under N-rich conditions. 

In our kMC simulations of N-polar GaN  $ (000\bar 1)$ growth  under N-rich conditions we  apply 1eV  barrier for  Ga surface diffusion  as evaluated in ab-initio calculations \cite{zywietz} and 1.8eV high barrier for diffusion of N adatoms. With  such a slow surface dynamics we model the crystal growth under large nitrogen flux at relatively high miscuts. We choose external fluxes and desorption rates in such a way that simulations follow  experimental crystal growth rates. As  a result the  experimentally observed  morphology patterns are well  reproduced. We explain that the surface instability appears due to the  too slow  diffusion of gallium atoms at relatively high flux of incoming atoms and because of the wide terraces as compared to the diffusion length of Ga adatoms\cite{misbah,politi,jeong,villain}. However if steps are located more densely,  crystal surface becomes  smoother under the same growth conditions.   In two component system of GaN, at miscut of 4$^o$, diffusion rate of N adatoms  is below  and of Ga adatoms close to the   limiting value for the domain formation on the terrace \cite{politi,jeong,villain,mode1,mode2,mode3_t}.    If the  diffusing particle has enough time to find a step edge  before the next one attaches to it the island nucleation process is suppressed.  When this condition is  fulfilled  we observe smooth  crystal growth realized either by step flow or  mixed made of step flow and  two dimensional nucleation process \cite{kallunki}.  It is interesting that we observe such growth mode for two component system, when one component is slower and the second moves with rate close to the expected limiting diffusion value. This limiting  value  can be decreased by  higher surface miscut, lower crystal growth rate or higher temperature. Our kMC simulations confirmed that all these methods lead to  lower surface roughness. Thus we can explain  experimentally observed differences in surface morphologies of N-polar GaN $(000\bar1)$  grown at $2^o$ and $4^o$ surface  miscuts.  However, due to heterogeneous growth and immobility of nitrogen adatoms the character of stable and unstable surface evolution is a  mixture  of step flow and 2-D or 3-D island growth \cite{jeong,villain,politi,mode3_t}. 

We find that low miscut angle $ (\le 2^o)$ results in characteristic surface instabilities (and subsequent high surface roughness), while for steeper cuts a smooth  double-step surface  pattern is observed. Other methods of surface smoothening lead to slightly different surface morphologies. For instance higher temperature smoothens the surface, but  contrary to the high miscut case no double step structure can be seen. Distances between all steps are equal suggesting that at higher temperature all steps move with the same rate at low  external particle flux (slow crystal growth rate). The resulting surface morphology is very similar to this for $4^o$ cut crystal. It can be seen when appropriate  correlation functions are compared. However there are  some differences in both patterns that suggest that dynamics of the process is slightly different.  

We have found that   step roughness  can be smoothened also  using  increased nitrogen  flux.  As nitrogen practically does not move along the surface at studied temperatures and the  rate  of gallium adatoms diffusion is also below limiting value, the character of this process is different than that studied  above.  The pattern that builds at the surface consists of islands that grow along  step-edges and finally attach to the step. Resulting  smooth crystal growth is realized by two dimensional island nucleation. 
We show that a flat surface is developed at relatively low miscut  at high nitrogen rates. We discuss numerical data as a function of the growth temperature, growth rate defined by Ga flux and substrate miscut angle.     The kMC results 
are compared with experimental data.

First in Section II the model is formulated, then in Section III we show how the surface structure changes with miscut, temperature, galium and nitrogen flux. Surface patterns obtained for the chosen parameters are  compared to the experimental data. Finally on the basis of the surface dependence on the nitrogen flux we consider possibility of the surface smoothening via use of higher  nitrogen  fluxes.

\section{Kinetic Monte Carlo model}
Numerical simulations of the system were performed with use of  two-component kMC model. It consists of gallium and nitrogen atoms arranged in  a crystallographic lattice of N-face  $(000\bar 1)$ of gallium nitride. Two species interact via nearest neighbor (NN) and next nearest neighbor (NNN) forces which form N-N, Ga-Ga, and Ga-N bonds.
Total energy of an adatom occupying one of the lattice sites is given by the sum over all NNs and NNNs surrounding it
\begin{equation}
\label{atom_en}
E_X=E_{GaN}\sum_{NN}n_i+E_{XX}\sum_{NNN}n_i
\end{equation}
where X corresponds to Ga or N atom and the value $n_i$ depends on the state of the neighboring site. If the site is occupied $n_i=1$ otherwise $n_i=0$. Energies associated with   interactions are $E_{NN}=0.3$, $E_{GaGa}=0.35eV$ and $E_{GaN}=1.6eV$ respectively. They were chosen in such a way that total energy of  Ga atom built in  the crystal  is 10.6 eV. This value is in agreement with molecular dynamic calculations performed by Wang et. al \cite{wang}  using a bond order potential.  

Single simulation consists of several Monte Carlo (MC) steps. Each of them starts with particle adsorption which occurs with probability equal to the external flux F.  Adsorbed particles are allowed to diffuse at the surface. That process is modeled in three stages. At first one of the neighboring sites is randomly chosen to be the target of a jumping particle. Next initial ($E_i$) and final ($E_f$) energies of jumping atom are calculated using eq. (\ref{atom_en}). In the third stage of diffusion process particles jump with probabilities:
\begin{equation}
\label{jump} 	
P_D=\left\{ \begin{array}{ll} 
\nu e^{\beta(E_f-E_i-{B^X_D})} & \textrm{when $E_i\geq E_f$}\nonumber \\
\nu e^{-\beta {B^X_D}} & \textrm{otherwise}
\end{array} \right.
\end{equation}
where $\beta=(k_BT)^{-1}$ and $\nu=10^{11}s^{-1}$ sets the time scale. $B_D^X$ denotes the diffusion barrier that is different for $X$=Ga, N adatoms.  It is known  that diffusion barriers for both components  under nitrogen rich conditions on  GaN$(000\bar1)$ surface are high\cite{zywietz,neugebauer,takeuchi,review}.  We use  barriers calculated in Ref.  \cite{zywietz}, and so Ga adatoms diffuse jumping over the  barrier $B_D^{Ga}=1eV$ and N adatoms jump  over barrier   $B_D^N=1.8eV$. According to the  Eq. \ref{jump} particle, which is strongly binded in its initial state, resides in a deep potential well and probability of the jump out  is low. However  when final state has low energy i.e. particle tries to move to the deep well the energy barrier between these states decreases and jump rate is high. Finally jumps  towards step from above occur  with the same probability like those far from the step in such  a way that effectively   no additional barrier for jumps  across step (Schwoebel barrier) is present in the studied systems.
During the last stage of a single MC step, the evaporation phenomenon is simulated. An assumption that incoming atoms can evaporate is necessary in order to reach equal number of Ga and N atoms within the growing crystal. We set the following   rate  of particle desorption:
\begin{equation}
D_X=\nu e^{-\beta(E_i+d_X)}
\end{equation}
where $d_X$ is the desorption potential, different for Ga and N adatoms. We set   $d_N=-0.7eV$ taking into account that  binding energy of single N atom at the surface $E_i$=5.7eV  while the same atom above the  $(000\bar1)$ GaN surface covered by nitrogen  $E_i=0.9eV$. This value of desorption barrier ensures that most of N adatoms that are not bonded to gallium atoms in the layer below, desorb from the surface, whereas the ones  that have  fallen onto the gallium layer stick to the surface.  For gallium we set  $d_{Ga}=0$. Such choice of desorption potentials provides that  the number  of adsorbed atoms of both  types equalizes  at the surface.  After evaporation stage new MC step begins and the above procedure starts from the beginning.

At the beginning of each simulated process, the system consists of $N_s$ straight steps. Helical boundary conditions in one direction and periodic in the second one are posed. Gallium nitride elementary cell consists of four mono-atomic layers, hence in order to close helical boundary condition properly, $N_s$ is always divisible by 4. 
Additionally,  because N-rich conditions are studied,  we  start the simulation with total N coverage of the surface, what changes during the growth process by evaporation of adatoms from the surface.

Summarize, the kMC model   used to describe N-polar  GaN($000\bar1$) growth under N-rich conditions we  build of two types of atoms:  Ga adatoms  that  move  slowly but are able to diffuse  across  terraces and reach  steps,   and  N adatoms that are almost immobile, but can  desorb from the surface and are frequently  adsorbed because of the relatively high flux of N atoms. The consequence of such assumptions is that both atomic monolayers (nitrogen and gallium) can grow smoothly within some range of parameters, but due to totally different  mechanisms. Gallium adatoms diffuse over the surface, and they start to build islands when incoming flux is too high, whereas nitrogen adatoms being practically immobile build islands all the time. Nitrogen islands have two different roles in the growth process: they build in the  step thus completing the  layer  formation, but  they also  change diffusion of Ga adatoms. 
 Below we analyze consequences of such adatom dynamics  for the  evolution of stepped surfaces.

\section{MBE experimental details}
\begin{figure} 
\includegraphics[width=7cm]{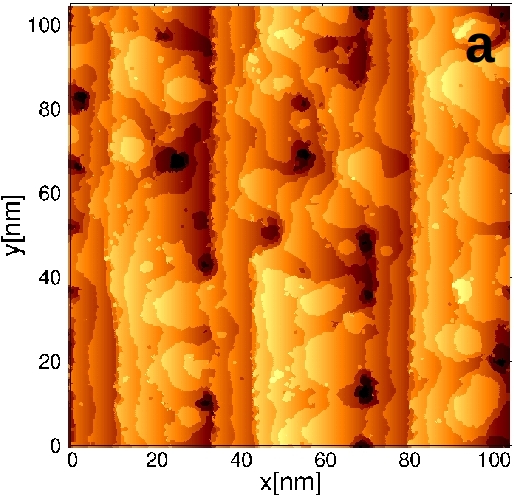} 
\includegraphics[width=8cm]{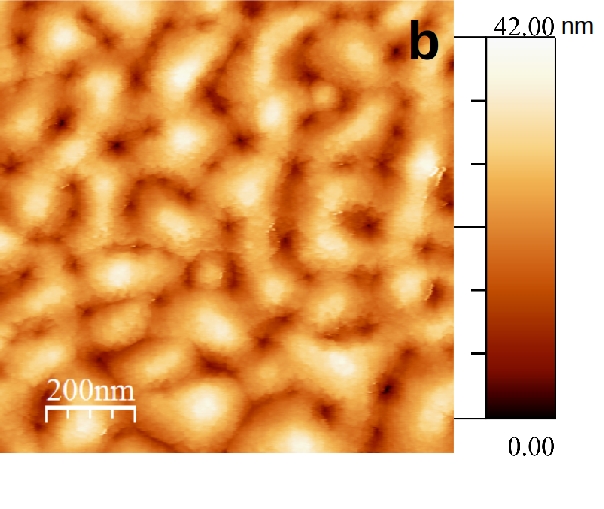} 
\caption{\label{eq} (color online) (a) Surface pattern of N-polar GaN$(000\bar1)$ layer simulated by kMC on $2^o$ miscut substrate. (b) Morphology of 200nm N-polar GaN layer grown by PAMBE under the condition used in the kMC simulations on $2^o$ miscut substrate.  Fluxes of atoms are $F_{Ga}=4 nm/min$ and $F_N=16 nm/min$ for both experiment and simulations. Temperature $T=750^oC$ in both pictures.}
\end{figure}
\begin{figure} 
\includegraphics[width=7cm]{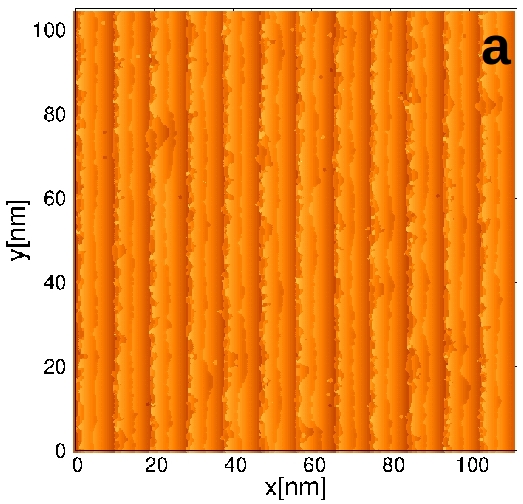} 
\includegraphics[width=8cm]{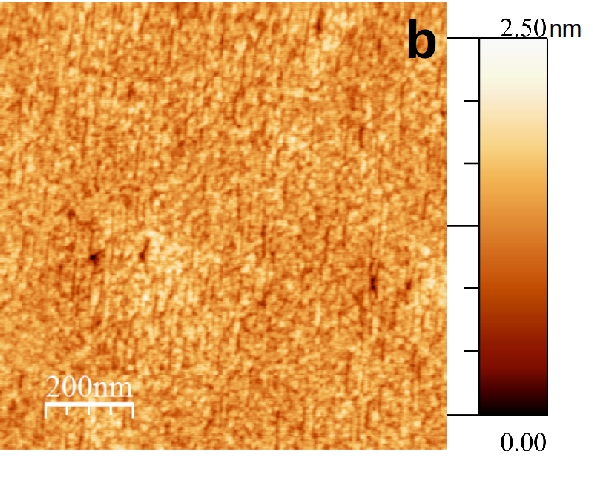}
\caption{\label{highMis} (color online) (a) Surface pattern of N-polar GaN$(000\bar1)$ layer simulated by kMC on $4^o$ miscut substrate. (b) Morphology of 200nm N-polar GaN layer grown by PAMBE under the condition used in the kMC simulations on $4^o$ miscut substrate.  Fluxes of atoms are $F_{Ga}=4 nm/min$ and $F_N=16 nm/min$ for both experiment and simulations. Temperature $T=750^oC$ in both pictures.} 
\end{figure}
 In parallel we carried out the growth of 200nm thick N-polar GaN layers by PAMBE
   custom design Gen20A MBE reactor were active nitrogen was supplied from Veeco RF plasma source. We used bulk, commercially available GaN substrates from Saint Gobain with threading dislocation density around 5 x 10$^7$ cm$^{-2}$. Prior to growth, substrates were mechanically polished to get miscut angle of $2^o$ and $4^o$ toward $[10\bar10]$ direction and chemomechanically polished to obtain atomically smooth surface. Epi-ready substrates were then mounted by gallium to single 2 inch wafer close to its center to ensure the same growth conditions at both crystals. Growth of 200 nm thick GaN was carried out at $750^{\circ}C$ out under nitrogen rich conditions using nitrogen and gallium fluxes of $F_N = 16nm/min$ and $F_{Ga} = 4nm/min$ (GaN equivalent growth rate) respectively. After the growth samples morphology was examined by atomic force microscope (AFM).

\section{Dependence of the surface ordering on the growth parameters}
\subsection{Miscut}
\ref{eq} presents the numerical (a) and experimental (b) results of N-polar GaN layer morphology for the case of $2^o$ miscut surface. For both cases (experimental and calculations) growth parameters like  temperature and atomic fluxes were  the same. The size of  calculated system is 105 nm $\times$ 105 nm, smaller than the area of AFM scan  which is $ 1 \mu $m $\times 1 \mu$m. Despite this  it can be seen that in both cases crystal does not grow smoothly.  In \ref{eq}a  several cavities can be easily seen, steps bend and become wavy leaving deep cavities on their way forward.
 In the process of further growth more profound structures  can build around them.
In \ref{eq}b we can see that   the surface is rough, dominated by islands. This is   typical  pattern expected for systems with slow adatom diffusion. 

The situation changes diametrically when growth is performed under the same conditions (impinging fluxes and temperature), but at the substrate cut $4^o$. \ref{highMis}a  presents the simulated surface pattern for $4^o$ miscut. Double-steps with relatively straight  edges are seen. \ref{highMis}b presents the AFM scan of N-polar GaN layer grown by PAMBE. We can see smooth, stepped  surface. The double step structure however is not so clearly visible  like in the simulated plot above. The double step structure is a consequence of  GaN  lattice geometry according to  which every second step perpendicular to [10$\bar1$0] direction has different bond arrangement.  Thus the  rate  of adatom adsorption at the  step  and resulting velocity  of every second step is different. 

In order to analyze and compare surface parameters more  precisely we have calculated  root mean square (RMS) roughness
\begin{equation}
\label{rms}
RMS=\sqrt{\frac{1}{N} \sum_i (h_i-h_0)^2}
\end{equation}
where $h_i$ is the z coordinate (height) of i-th site over leveled surface and  $h_0$ is mean system height. RMS parameters for numerical results presented in \ref{eq}a and \ref{highMis}a are   0.38nm and  0.15nm respectively. The same parameters calculated for corresponding experimental data (\ref{eq}b and \ref{highMis}b)  are 6.11nm and 0.49nm.  Roughness measured for experimental surfaces is obviously higher due to the surface distortion  or local defects, however one can notice that RMS parameters for lower miscut are higher both in the experiment and simulations as well. Numerically obtained  surfaces are smoother also due to the shorter time of system evolution, which for experiments was about 40 minutes and 2-3 minutes for simulations as well as due to different size of the surface area.
\begin{figure}
\includegraphics[width=9cm]{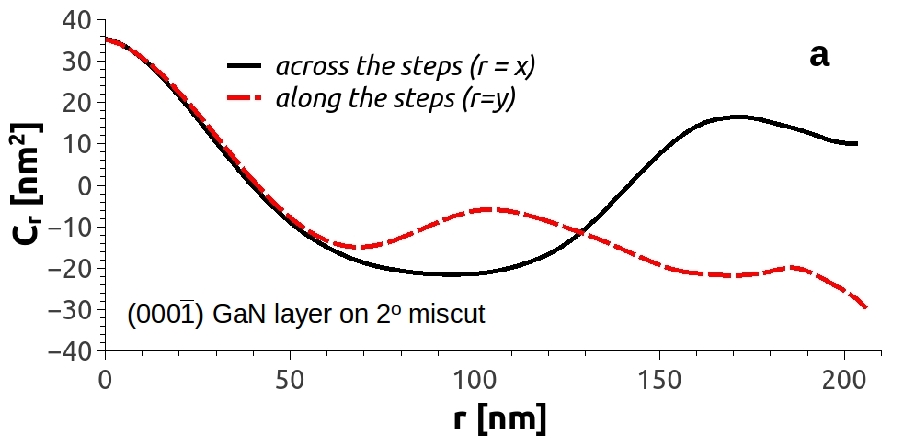}
\includegraphics[width=9cm]{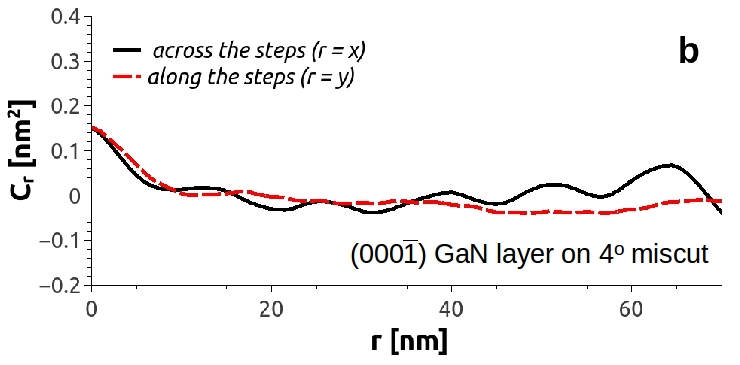}
\caption{\label{xx} (color online) Correlation functions for experimental   surfaces  for a) $2^o$ miscut  and b) $4^o$ miscut.  The other parameters like in \ref{eq} and \ref{highMis}. Solid lines denote correlation calculated perpendicularly to  to the initial step orientation - along x axis, and dashed lines denote correlations parallel to  steps - along y axis. }
\end{figure}
\begin{figure}
\includegraphics[width=9cm]{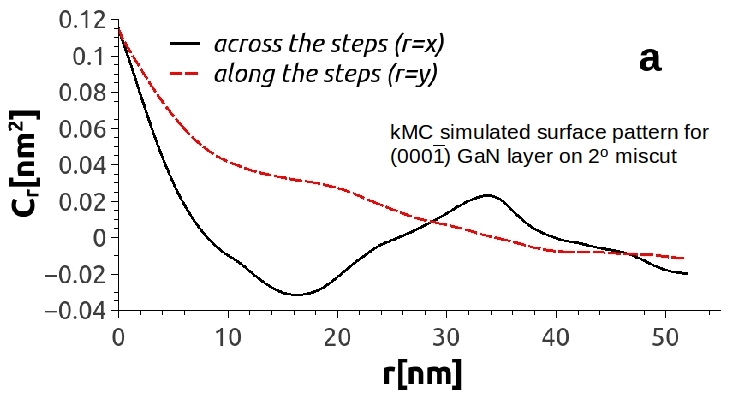}
\includegraphics[width=9cm] {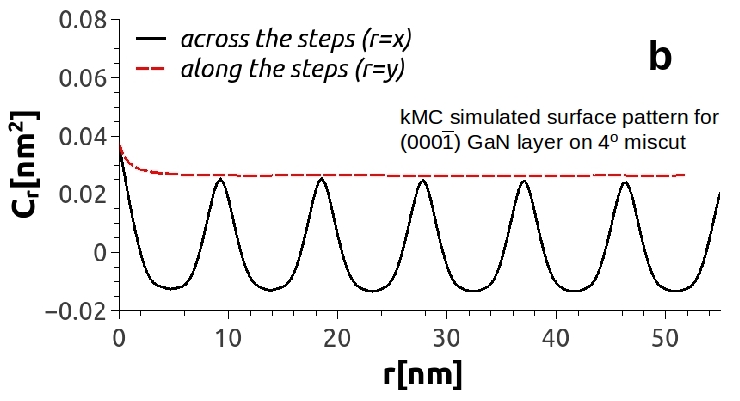}
\caption{\label{corr} (color online) Correlation functions for simulated  surfaces  for a) $2^o$ miscut and  b) $4^o$ miscut.  The other parameters like in \ref{eq} and \ref{highMis} . Solid line denotes correlation calculated perpendicularly to  to the initial step orientation - along x axis, and dashed line denotes correlations parallel to  steps - along y axis. }
\end{figure}
\begin{figure} \includegraphics[width=12cm]{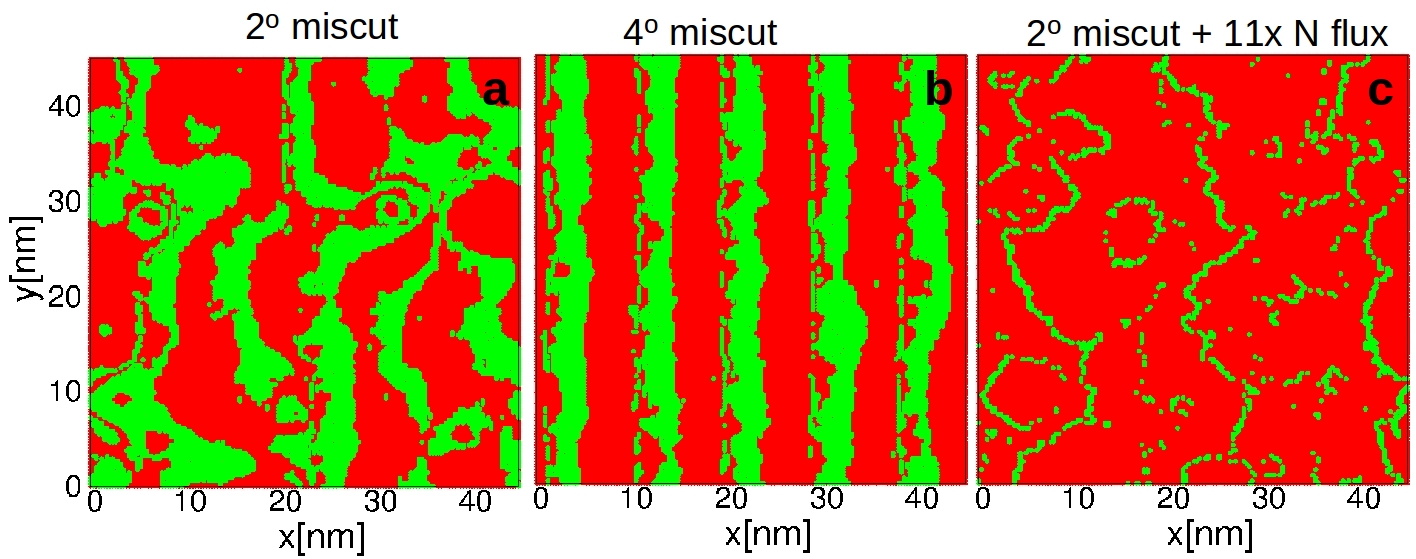} 
\caption{\label{zoom} (color online) a) Simulated pattern of  $2^o$ miscut from Fig 1a. b) $4^o$ miscut from Fig 2a and c) $2^o$ miscut at high $N$ flux $F_N=$180nm/min and other parameters  the same as in a). Red (dark gray) points denote nitrogen and green (light gray) gallium atoms.}
\end{figure}

In order to illustrate and analyze surface features, in particular their periodicity and amplitude we calculated   correlation functions along  and across steps according to the formula:
\begin{equation} 
C(\bar{r_i}-\bar{r_j})=\sum_i [h(\bar{r_i})-h_0][h(\bar{r_j})-h_0]
\label{corr_eq}
\end{equation}
where the difference in position  $\bar {r_i}-\bar{r_j}$ is calculated along x or y axis i.e. across ($r=x$) or along  ($r=y$) steps. \ref{xx} presents the correlation functions 
calculated for experimentally measured surface morphology of N-polar GaN$(000\bar{1})$ layers grown by PAMBE on (a)  $2^o$ miscut  and  (b) $4^o$ micut GaN substrate.  We can see that the scale of variability of the $2^o$ curve is much larger than that for  $4^o$ surface i.e. $\pm 30 nm^2$ vs. $\pm 0.2nm^2$ respectively. 
\ref{xx}a. shows that the correlation function calculated for the surface  of $2^o$   oscillates in both directions. The  length  scale of its changes  along and across steps is 
much longer than  bilayer step  distance ($ \approx$20nm).  We can see that the characteristic correlation length  is shorter   across  than   along steps.
 The   correlation function across steps of $4^o$ miscut surface (\ref{xx}b)  shows  periodicity of  around 10 nm length  denoting  bi-step surface structure.  The function along step is flat and close to zero describing smooth surface. 

Correlation functions for simulated $(000\bar1)$ GaN surfaces on $2^o$ miscut   (\ref{corr}a) and 4$^o$ miscut (\ref{corr}b) exhibit many similarities to the experimentally corresponding counterparts.
The  characteristic  length scales at $2^o$ cut are  shorter in both directions, thought it reflects less corrugated structure due to short available simulation times. Nevertheless the period of correlation function oscillations is   still longer than bilayer step distance. Correlation function  across the steps for $4^o$ miscut $(000\bar1)$, presented in \ref{corr}b,  clearly shows  the step structure.  Oscillations of the function   are very regular and all with the  same amplitude what  corresponds to the very regular structure. Thus the character of  surface morphology is very well reflected by the shape of correlation function.

The analysis of the correlation function for the $(000\bar1)$ GaN surface morphologies obtained on $2^o$ miscut and $4^o$ miscut substrates confirms  that   the surface of a crystal grown at substrate with higher miscut is more regular then the surface which evolved at the substrate with lower miscut. In order to understand the causes of observed differences we carry out the following analysis: 
Formation of Ga 2D islands on the terraces is probable for $2^o$ miscut because the average diffusion length for Ga adatoms  until they meet next particle is around 8nm, which is close to the terrace width for 2$^o$ cut.
To be more precise we  calculate the limit terrace length for the domain formation \cite{politi,villain,mode3_t,mode1,mode2}, that is given by  the relation $l_{cr}=(D/F)^{1/6}$. For our data, due to the low diffusion barrier, the limit terrace length   is around  $l_{cr}^N=4a$ for nitrogen adatoms and $l_{cr}^{Ga}=17a$ for gallium adatoms, where $a$ denotes GaN lattice constant $a=0.319nm$, This gives $l_{cr}^N=1.3nm$ and $l_{cr}^{Ga}=5.44nm$ as compared with the terrace width $l=8a$ at the surface of $4^o$ miscut and  $l=16a$ in the $2^o$ case. This means that in this  for $2^o$ miscut $(000\bar1)$ GaN surface  nitrogen adatoms have tendency of creating islands, whereas gallium adatoms are at the crossover length. However at the temperatures studied the surface  has a  tendency to build bi-layer structures. Hence terrace width that has to be considered as being  two times wider.  As a result adatoms at  $4^o$ miscut surface are already within the crossover region.

In such situation, particles adsorbed at the steeper  surface start to form islands, but because these islands  built up  short distance to step edges. Steps move forward due to the attachment of other particles. In such a way islands are absorbed at the steps . It can be seen in \ref{zoom}b. Green points that are plotted for gallium atoms  are arranged in pattern that looks like many islands glued together. Nitrogen adatoms seem to have not much importance here, because there are enough of them all around. Hence the system   forms steps straight and distributed regularly in space. On the other hand, when miscut angle is low, terraces are wider and particles need more time to reach the step. In such  case the probability of  2-dimensional nucleation is higher. As a result of this process islands would  appear at terraces. They are partly adsorbed by steps, leaving some cavities beside them and  irregular wavy pattern builds up (\ref{zoom}a).  Less mobile nitrogen adatoms  that are adsorbed and desorbed  seem to  play a  role in building  3D islands. On building next layer on top of island they can be source of effective Schwoebel barrier for wandering gallium adatoms. For much higher nitrogen flux this reasoning does not work. Nitrogen adatoms are everywhere and they are rather smoothening the surface of growing $2^o$ crystal as we can see in \ref{zoom}c.
\begin{figure} \includegraphics[width=10cm]{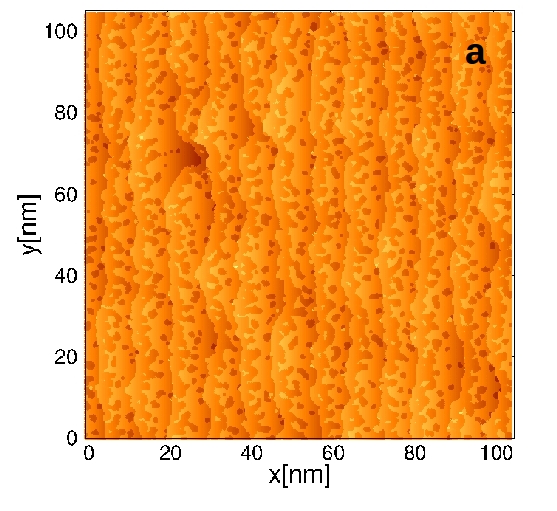}
 \includegraphics[width=10cm]{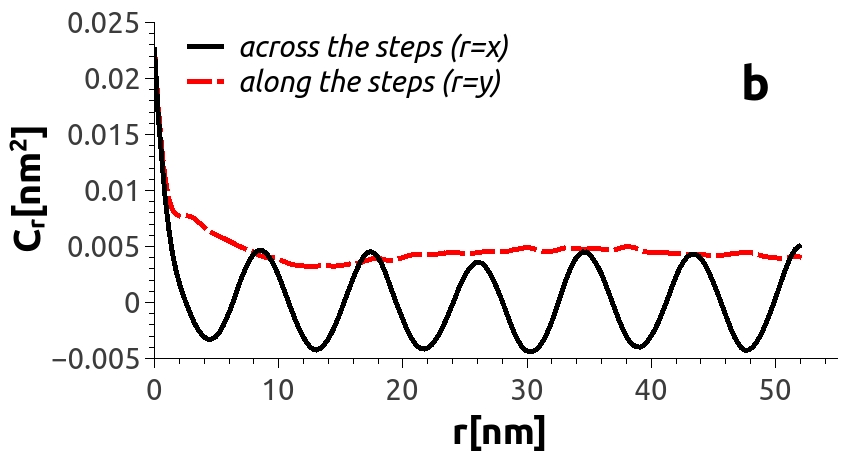}
\caption{\label{HT} (color online) (a) Surface pattern of N-polar GaN$(000\bar1)$ layer simulated by MC on $2^o$ miscut substrate at high $T$. $F_{Ga}=4n$m/min,$F_N=16n$m/min, $T=850^oC$ . (b) Correlation functions  calculated for this surface. Solid line denotes correlation calculated perpendicularly to  to the initial step orientation - along x axis, and dashed line denotes correlations parallel to  steps - along y axis. }
\end{figure}
\begin{figure} \includegraphics[width=10cm]{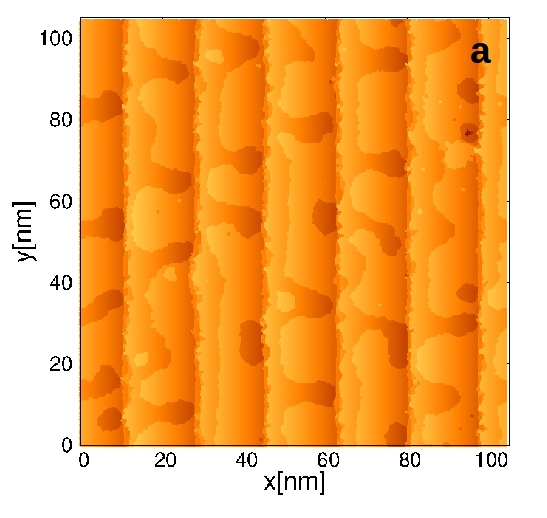} 
\includegraphics[width=10cm]{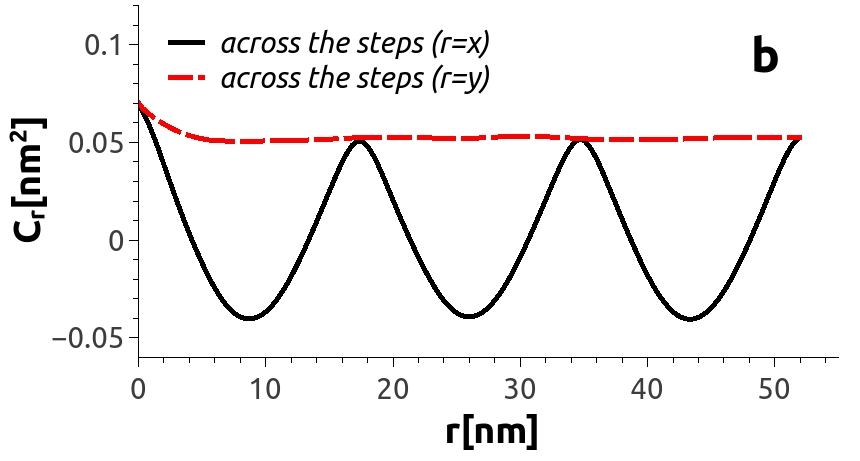}
\caption{\label{LF} (color online)) (a) Surface pattern of N-polar GaN$(000\bar1)$ layer simulated by kMC on $2^o$ miscut substrate for low F. $F_{Ga}=1n$m/min, $F_N=4n$m/min, $T=750^oC$. (b) Correlation functions  calculated for this surface. Solid line denotes correlation calculated perpendicularly to  to the initial step orientation - along x axis, and dashed line denotes correlations parallel to  steps - along y axis. }
\end{figure}
\begin{figure} \includegraphics[width=10cm]{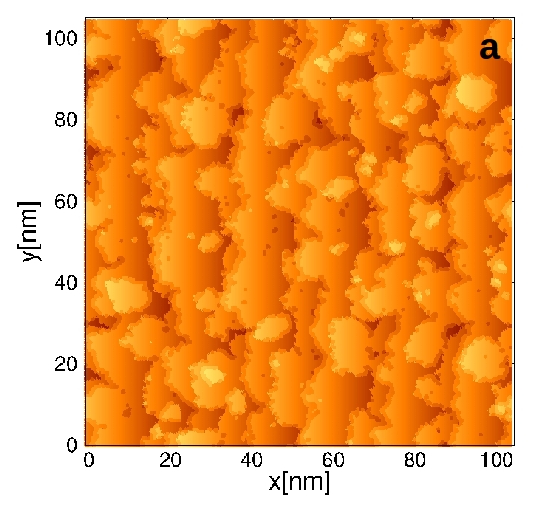} 
\includegraphics[width=10cm]{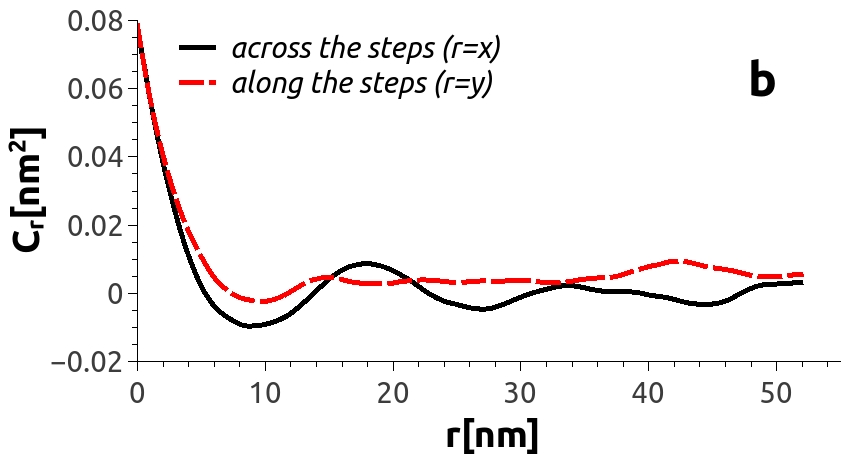}
\caption{\label{N_flux} (color online) (a) Surface pattern of N-polar GaN$(000\bar1)$ layer simulated by kMC on $2^o$ miscut substrate for high N flux. $F_{Ga}=4n$m/min, $F_N=180n$m/min, $T=750^oC$. (b) Correlation functions  calculated for this surface. Solid line denotes correlation calculated perpendicularly to  to the initial step orientation - along x axis, and dashed line denotes correlations parallel to  steps - along y axis. }
\end{figure}
The surface morphologies of GaN $(000\bar1)$ layers grown on $2^o$ miscut (\ref{eq}) and on $4^o$ miscut \ref{highMis}  substrates clearly evolve in two directions. From the microscopic  point of view, there is a significant  difference  between  the time that adatoms need to reach atomic steps  and the probability for encountering one another.
\subsection{Temperature}
 Higher miscut successfully flattens the surface by shortening the time that  adatom needs to reach the step.  Knowing that we carried out another simulation for $2^o$ miscut applying a higher growth  temperature, thus increasing surface  diffusion  rate to check if this will result in smoother surface.
In \ref{HT} we show simulated surface morphology for $(000\bar1)$ GaN layer on $2^o$ miscut surface obtained at the temperature $850^oC$, that is  
 $100^oC$  higher than the reference  (cf. Fig 1). 
 In such case diffusion is 4 times faster than in the reference  example. It should compensate 4 times longer lasting  diffusive walk, what means effectively 2 times longer terrace width. Comparing the surface morphologies in Fig.  1a and Fig. 6a we see the positive effect  of raised temperature on the surface smoothness. Calculated RMS value for the N-polar $(000\bar1)$ GaN surface simulated at 850$^oC$ is $0.15 nm$, what is low value. However the morphologies obtained at raised temperature on $2^o$ miscut $(000\bar1)$ substrate (\ref{HT}) and at reference temperature on $4^o$ miscut $(000\bar1)$ GaN substrate are different. As can be concluded from the correlation functions, plotted in Fig. 4b and \ref{HT}b. Regular pattern across steps is clearly visible, but with no sign of double step structure.  In \ref{HT}b the periodicity of the correlation function is around 10nm across the steps, while  along the 
 steps nice smooth  terraces are indicated   by low, close to zero correlation  line. Such structure can be also seen in Fig. 4b. However, when we look closer at the surface morphology in Fig. \ref{HT} we can see  that every  second step has irregular  structure similar  to that seen in Fig. \ref{zoom}b. More detailed study shows that these island-like structures are build by nitrogen and not by gallium atoms.  Gallium  steps at the other side have tendency to bend at characteristic length related to the diffusive distance at given temperature. 

\subsection{Growth rate}
The other method of surface smoothening  is to use a  lower flux $F$ of incoming particles i.e. lower growth rate. In such a way  diffusing adatom has more time to   wander around on  the surface and as a result  the steps should become more straight. In \ref{LF}a we see an effect of the slow growth rate  (flux are 4 times slower than in the reference surface in \ref{eq}). Again we can see a rather smooth surface. It differs from this obtained by increased temperature. We can see bi-step structure, like in the case of  $4^o$ miscut. Such structure is reflected in the shape of the correlation functions presented in \ref{LF}b. Correlation in the direction perpendicular to steps is very similar to this in \ref{highMis}, but in two times larger distance scale. At the same time correlation along step quickly decreases to zero.  $RMS=0.27nm$ is  higher than in the  case of surface of crystal grown at higher temperature \ref{HT}.

\subsection{N-excess}
One more simulation was carried out to study the effect of the increased  nitrogen flux on surface morpholog. 
 In \ref{N_flux}a we show the surface of GaN$(000\bar1)$ on  $2^o$ miscut obtained under the growth conditions of high N excess, i.e. the flux of incoming nitrogen is 30 times higher
 than gallium flux. Part of this surface was already presented above in \ref{zoom}c.
It can be seen that the surface is much smoother that this in \ref{highMis}, however not in the same manner as $4^o$ miscut case as seen in the simulated system in \ref{highMis}a.  
 It is easy to note that the increase of nitrogen amount in the system leads to the improvement of surface quality.  There  are no cavities seen in \ref{N_flux} and the  height differences are not so large as in the reference  $(000\bar1)$ GaN surface.
However,  the overall surface structure is not so regular as in the cases studied before.
  The calculated value  of  $RMS=0.28nm$ is lower than the reference $2^o$ miscut GaN $(000\bar1)$ layer.
 When we look at the correlation function along steps presented  in \ref{N_flux}b we see that  it decreases very quickly, like for smooth structures. The correlation function  across the steps shows oscillations of characteristic length  for double step structure i.e. 20nm. 
\begin{figure}
\includegraphics[width=9cm]{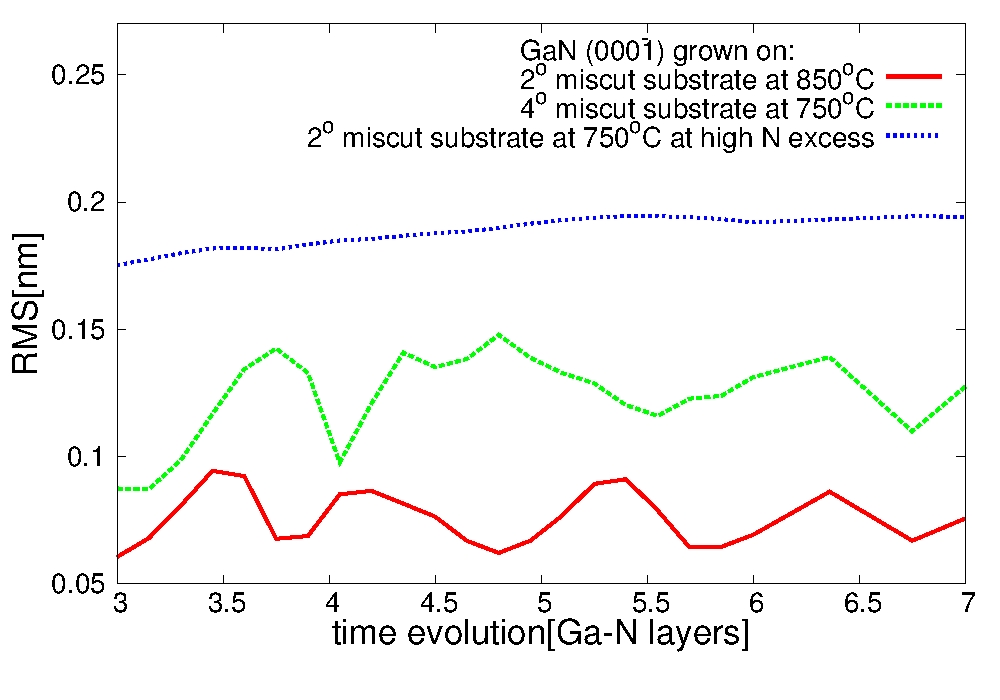}
\caption{\label{RMS} (color online) RMS as  a function of coverage shows periodicity of the process for the system of $4^o$ cut and for high temperature growth for $2^o$ cut compared to steady evolution for high nitrogen growth.}
\end{figure}
In the last part of our study we analyzed the evolution of surface RMS parameter as a function of surface coverage, counted in the numbers of grown layers for three  cases of $(000\bar1)$ GaN growth: high miscut, high temperature, and high N flux. The result is shown 
 in Fig.~\ref{RMS}.  
For  the case of layer grown at the  $4^o$ miscut GaN substrate  and high temperature  $2^o$ miscut growth process is evidently periodic.  One can see oscillations in RMS when the growth proceeds. It can be interpreted as layer by layer growth via creation of  islands, than absorption of these islands by approaching steps, and the cycle ends up by next layer to be completed \cite{l_b_l,l_b_l2}. We can clearly see periodicity of the length of one monolayer in the plotted curves. Typically, the higher temperature process is the smoother line  we would expect, but it has the same period  related to the number of layers. Third curve is plotted for high N-excess case RMS value  stays at the same, relatively high level, showing that  the growth  process is realized   in a different way. We attribute it to the island formation and coalescence  that results in a stable RMS value over the growth time.

\section{Conclusions}
Kinetic Monte Carlo simulations have been employed to study the surface evolution of N-polar  GaN$(000\bar1)$ layers grown on the substrates of $2^o$ and $4^o$ miscut. The numerical results are compared with experimentally observed surface morphologies of GaN layers grown by MBE.
Due to the high diffusion  barriers  during the growth of $( 000\bar1)$ GaN under N-rich conditions the resulting surface is irregular and  rough. 
Such surface can be observed for the  crystal  grown at  $2^o$ surface miscut at temperature around 750$^o$C.  Four  different methods of smoothening the surface were investigated using kinetic Monte Carlo method: (a) growth of GaN$(000\bar1)$ on high misut substrates, (b) at higher temperature, (c) at low growth rate and (d) using  high N-excess. All of them lead to smoother crystal  surfaces.   The surface of $4^o$ cut seems to be the most regular one, that has been also confirmed by  by experimental results. Other approaches like: higher temperature or lower growth rate have also work as smoothening factor, but the character of  the surface morphology is different than this for the high miscut surface.  Their further evolution seems to happen  less smoothly due to meandering steps and nucleation on terraces that lead to the formation of cavities.  The surface that emerged during the growth at high N-excess using  very high nitrogen flux is quite unusual, because microscopically  it seems to be  irregular, but correlation functions  exhibit rather  regular structure. Thus, we  conclude  that under high N-excess GaN$(000\bar1)$ growth  happens regularly and the surface morphology remains smooth due to stable island formation and coalescence on the terraces.  Therefore the use of high nitrogen flux for the growth of GaN$(000\bar1)$ under N-rich conditions seems to be a good method to obtain smooth surface morphology.

\begin{acknowledgement}
Research supported  by the National Science Centre(NCN) of Poland
(Grant NCN No. 2013/11/D/ST3/02700)

\end{acknowledgement}


\begin{thebibliography}{99}
\bibitem{rajan} S. Rajan, A. Chini, M. H. Wong, J. S. Speck, and U. M. Mishra, J. Appl. Phys., 102, 044501, 2007.
\bibitem{hoi}W. M. Hoi, K. Stacia, S. D. Nidhi, D. D. J., K. Seshadri, F. B. David, L. Jing, F. N. A., A. Elaheh, S. Uttam, C. Alessandro, R. Siddharth, P. D. Steven, S. James, M. K. Umesh, Semiconductor Science and Technology, 28 074009 (2013).
\bibitem{keller}S. Keller, H. Li, M. Laurent, Y. Hu, N. Pfaff, J. Lu, D. F. Brown, N. A. Fichtenbaum, J. S. Speck, S. P. DenBaars, U. K. Mishra, , Semiconductor Science and Technology, 29 113001 (2014).
\bibitem{zywietz} T. Zywietz, J. Neugebauer, M. Scheffler Appl. Phys. Lett. 73, 487 (1998).
\bibitem{neugebauer} J. Neugebauer et al., Physical Review Letters 90, 056101 (2003).
\bibitem{takeuchi} N. Takeuchi, A. Selloni, T. H. Myers, and A. Doolittle, Phys. Rev B 72,115307 (2005). 
\bibitem{review} R. M. Feenstra, J. E. Northrup and J. Neugebauer, MRS Internet J. Nitride Semicond. Res. 7, 3 (2002). 
\bibitem{monroy}E. Monroy, E. Sarigiannidou, F. Fossard, N. Gogneau, E. Bellet-Amalric, J.-L. Rouvi{\`e}re, S. Monnoye, H. Mank, B. Daudin,  Applied Physics Letters, 84 3684 (2004).
\bibitem{cheze} C. Ch{\`e}ze, M. Sawicka, M. Siekacz, H. Turski, G. Cywinski, J. Smalc-Koziorowska, J. L. Weyher, M. Kry{\'s}ko, B. Lucznik, M. Bockowski, C. Skierbiszewski et al., Applied Physics Letters 103, 071601 (2013).
\bibitem{zauner} A. R. A. Zauner, J. L. Weyher, M. Plomp, V. Kirilyuk, I. Grzegory, W. J. P. van Enckevort, J. J. Schermer, P. R. Hageman, P. K. Larsen, Journal of Crystal Growth, 210 435 (2000).
\bibitem{krzyz1} M. A. Za{\l}uska-Kotur, F. Krzy{\.z}ewski, S. Krukowski, Micha{\l} Leszczynski, Robert Czernecki  Cryst. Growth \and Design 13, 1006 (2013). 
\bibitem{krzyz2} M. A. Za{\l}uska-Kotur, F. Krzy{\.z}ewski, S. Krukowski, Journal of  Applied  Physics, 109  023515 (2011).
\bibitem{krzyz3} M. A. Za{\l}uska-Kotur, F. Krzy{\.z}ewski,   Journal of Applied Physics 111, 114311 (2012).
\bibitem{krzyz4} M. A. Za{\l}uska-Kotur, F. Krzy{\.z}ewski, S. Krukowski, J. Cryst. Growth 343 138 (2012).
\bibitem{misbah} C. Misbah, O. Pierre-Louis, and Y. Saito, Rev. Mod. Phys. 82, 981 (2010).
\bibitem{jeong} H.-C. Jeong, E. D. Williams, Surf. Sci. Rep. 34, 171 (1999)
\bibitem{politi}P. Politi, G. Grenet, A. Marty, A. Ponchet, J. Villain,Physics Reports 324 (2000) 271
\bibitem{villain}Jacques Villain ,Alberto Pimpinelli, Leihan Tang and Dietrich Wolf, J. Phys. I France 2 2107 (1992).
\bibitem{mode3_t}J. W. Evans, P. A. Thiel, M. C. Bartelt, Surface Science Reports 61, 1 (2006).
\bibitem{mode1}G. B. Stephenson, J. A. Eastman, C. Thompson, O. Auciello, and L. J. Thompson, Appl. Phys. Lett., 74, 3326 (1999).
\bibitem{mode2}Edith Perret, M. J. Highland, G. B. Stephenson, S. K. Streiffer, P. Zapol, P. H. Fuoss, A. Munkholm, and C. Thompson, Applied Physics Letters 105, 051602 (2014).
\bibitem{kallunki}M. Rusanen, I. T. Koponen, and J. Kallunki, Eur. Phys. J. B, 36, 141-147 (2003).
\bibitem{wang} K. Wang, J. Singh, D. Pavlidis J. Appl. Phys. 76, 3502 (1994).
\bibitem{l_b_l}R. Kunkel, B. Poelsema, L.K. Verheij, G. Comsa, Phys.
Rev. Lett. 65, 733 (1990) 
\bibitem{l_b_l2} P. {\^S}milauer, M. R. Wilby, D. D. Vvedensky, Phys. Rev. B
47, 4119 (1993)
\end{thebibliography}
\end{document}